\documentclass[aps,scriptaddress,twocolumn,prx,showkeys,amsmath,amssymb,nofootinbib, preprintnumbers,notitlepage,10pt]{revtex4-1}

	\usepackage{amsmath}
	\usepackage{makeidx}
	\usepackage{amsfonts}
	\usepackage[ansinew]{inputenc}
	\usepackage[usenames,dvipsnames]{pstricks}
	\usepackage{subfigure}
	\usepackage{epsfig}
	\usepackage[colorlinks,hyperindex]{hyperref}
	\usepackage{soul} 

	\newcommand{\beq}{\begin{equation}}
\newcommand{\eeq}{\end{equation}}
\newcommand{\bea}{\begin{eqnarray}}
\newcommand{\eea}{\end{eqnarray}}

\newcommand{\Br}[1]{\ensuremath{\left ( #1\right)}}
\newcommand{\Sq}[1]{\ensuremath{\left [#1\right]}}

\newcommand{\nl}{\nonumber \\ }

\newcommand{\DD}[2]{\ensuremath{\frac{\mathrm{d} #1}{\mathrm{d} #2}}}

\renewcommand{\i}{\ensuremath {\mathrm{i}}}

\newcommand{\sigmaz}{\ensuremath{\sigma_{\textrm{z}}}}

\newcommand{\subrm}[1]{\ensuremath{_\mathrm{#1}}}

\makeindex

\begin{document}
\title{Detecting gravitational decoherence with clocks: Limits on temporal resolution from a classical channel model of gravity}

\author{Kiran Khosla$^{1,2}$, Natacha Altamirano$^{3,4}$}
\email{k.khosla@uq.edu.au, \\naltamirano@perimeterinstitute.ca}
\affiliation{$^{1}$Center for Engineered Quantum Systems, The University of Queensland, Brisbane, QLD, 4067, Australia.}
\affiliation{$^{2}$Department of Physics, The University of Queensland, Brisbane, QLD, 4067, Australia.}
\affiliation{$^{3}$Perimeter Institute for Theoretical Physics, 31 Caroline St. N., Waterloo, ON, N2L 2Y5, Canada}
\affiliation{$^{4}$Department of Physics and Astronomy, University of Waterloo, ON, N2L 3G1, Canada}

\date{\today} 

\begin{abstract}
The notion of time is given a different footing in Quantum Mechanics and General Relativity, treated as a parameter in the former and being an observer dependent property in the later. From a operational point of view time is simply the correlation between a system and a clock, where an idealized clock can be modelled as a two level systems. We investigate the dynamics of clocks interacting gravitationally by treating the gravitational interaction as a classical information channel. In particular, we focus on the decoherence rates and temporal resolution of arrays of $N$ clocks  showing how  the minimum dephasing rate scales with $N$, and the spatial configuration. Furthermore, we consider the gravitational redshift between a clock and massive particle and show that a classical channel model of gravity predicts a finite dephasing rate from the non-local interaction. In our model we obtain a fundamental limitation in time accuracy that is intrinsic to each clock.  \end{abstract}


\maketitle

\textit{Introduction} --- Despite the success of General Relativity (GR) and Quantum Mechanics (QM) to describe nature at large and small scales respectively there is still an open question as to what the interplay is between these two theories. This question becomes fundamental when treating the nature of time operationally, specifically when considering how an observer measures time in GR and QM. Operationally, a clock is a reference and the notion of time emerges as a correlation between the clock and a system~\cite{gambini_free_2009,page_evolution_1983}. Even with a fundamental flow of time, any observer limited to only measurements of quantum systems will not be able to access this fundamental flow \cite{downes_optimal_2011} with zero uncertainty.

Recently, Castro {\it{et.al.}}~\cite{castro-ruiz_entanglement_2015} proposed a physically motivated quantum mechanism that produces fundamental uncertainty in measurements of coupled two level systems (clocks). The key idea in their model is that the mass energy equivalence in quantum clocks leads to a Newtonian coupling between them. This interaction entangles the clock states, and therefore a measurement of any single clock necessarily decoheres distant clocks, limiting the temporal resolution of distant observers. In this case the decoherence is entirely a consequence of mass energy equivalence with unitary quantum mechanics, similar to Ref.~\cite{pikovski_universal_2015}. In this letter we take a different approach by treating the gravitational interaction between  clocks in the context of classical channel gravity (CCG): a recent proposal that treats gravity as a fundamentally classical interaction~\cite{kafri_classical_2014}.
This model describes the gravitational interaction between quantum systems and results in noisy dynamics with decoherence rates similar those predicted by Di\'{o}si~\cite{diosi_universal_1987,diosi_models_1989} and Penrose~\cite{penrose_gravitys_1996}. The unitary quantum interaction considered in Ref.~\cite{castro-ruiz_entanglement_2015} is replaced by the master equation derived in Ref.~\cite{kafri_classical_2014}, resulting in non-unitary dynamics for all particles that interact gravitationally. We will show that the key difference between the two proposals resides in the ability of the gravitational interaction to entangle the clocks: in Ref.~\cite{castro-ruiz_entanglement_2015} the decoherence is a result of tracing out parts of an entangled state generated by standard unitary quantum mechanics, whereas in our model the decoherence is a consequence of the postulated quantum-classical interaction. Consequently, the limited temporal resolution is fundamental to each clock and we will discuss this in the context of operational time. There are several proposals to probe relativistic behaviour of quantum mechanics in the lab~\cite{pikovski_universal_2015,zych_quantum_2011,schmole_micromechanical_2016,lin_radiation_2016}, which focus on including standard principles of relativity within the framework of quantum mechanics. However, since CCG is fundamentally a modification of the equations of motion for quantum systems interacting gravitationally, we focus on potentially detectable deviations from standard quantum mechanics~\cite{pikovski_probing_2012,li_detecting_2016} in a post-Newtonian regime -- by allowing energy-mass equivalence.

This letter is organized as follows. Firstly we show that the master equation derived in CCG results in a fundamental phase diffusion for spin $\frac12$ systems, and the coherence time --- inverse dephasing rate --- is given by the gravitational interaction rate. We then extend the model to consider multiple spin $\frac12$ systems, and characterize how the dephasing rate depends on the number of clocks, as well as their geometric arrangement comparing our results with current experiments. Finally we show that CCG implies a non-zero dephasing in spin $\frac12$ clocks from earth's gravitational field. We conclude with a discussion of the implications of our model and its testability.  \\


\textit{Coupled clocks}\label{sec:clocks} --- In the following we consider a two level system clock with its spin processing around the z-axes of the Bloch sphere~\cite{ludlow_optical_2015}. The free clock Hamiltonian is $H = \hbar \omega \sigma_z$ where $\omega$ is the clock frequency and $\sigma_z$ is the Pauli-$z$ matrix. From Einstein's mass energy equivalence the clock has an effective mass $m = m_0 + H/c^2$ where $m_0$ is the rest mass of the clock and $c$ the speed of light. Note that this mass operator does not violate Bargmann's super selection rules~\cite{zych_quantum_2015,zych_the}, and is in a similar spirit to Refs.~\cite{zych_general_2016,zych_general_2012}. From the quantum correction to the mass, two clocks with rest masses $m_1$ and $m_2$, separated by a distance $d_{12}$ experience a Newtonian interaction 
\begin{eqnarray}
H_I\! &=&\! -\frac{G m_1m_2}{d_{12}} - \frac{G\hbar}{d_{12}c^2}\Br{m_2\omega_{1}\sigma_z^{(1)} +m_1\omega_2\sigma_z^{(2)}} \nl
\!\!& &\!- \frac{G\hbar^2 \omega_1\omega_2}{d_{12}c^4}\sigma_z^{(1)}\sigma_z^{(2)}\,,
\label{eq:HI}
\end{eqnarray}
that couples their internal energy states. The first term in Eq.~\eqref{eq:HI} is a constant potential, and the second term is the gravitational redshift on clock $1$ (clock $2$) from the rest mass of clock $2$ (clock $1$) which can be absorbed into the frequencies $\omega_{1,(2)}$ and therefore both terms are neglected. The last term is a coherent quantum interaction between the clocks that arises from the mass energy equivalence. We now examine this non-local gravitational interaction as if it were mediated by a classical information channel. That is, only classical information can be exchanged between the two separated quantum systems in a way that preserves the interaction Hamiltonian. Kafri {\it{et.al.}}~\cite{kafri_classical_2014} constructed the classical channel model where the Newtonian gravitational interaction between two quantum observables $\hat{x}$ and $\hat{y}$ emerge as a measurement and feedback process. The operators $\hat{x}$ and $\hat{y}$ are both measured with results $\bar{x}$ and $\bar{y}$ respectively~\cite{jacobs_straightforward_2006}, and a feedback control Hamiltonian $H_{fb} = \hbar g(\bar{x}\hat{y}+ \hat{x}\bar{y})$ replaces the unitary evolution generated by $H_I = \hbar g\hat{x}\hat{y}$. The net result of this  process is to preserve the systematic dynamics generated by the interaction Hamiltonian $H_I$~\cite{wiseman2009quantum}. However, the measurement and feedback process leads to dissipative evolution that cannot be avoided. This interaction is explicitly a local operation and classical communication (LOCC) process and therefore cannot lead to entanglement\cite{bennett_purification_1996,nielsen_conditions_1999}. However, note that non-local entanglement is still possible through other quantum interactions present in the system such us the Coulomb interaction. Even though CCG can be considered in the framework of quantum measurement and control, this is only a convenient mathematical tool to derive a master equation, and CCG does not require the existence of an agent to perform any such measurements or feedback: the decoherence is a natural consequence of coupling quantum and classical degrees of freedom as considered in~\cite{kafri_noise_2013,kafri_bounds_2015}, without violating quantum state positivity and Heisenberg's uncertainty principle~\cite{diosi_hybrid_2014,diosi_quantum_1995}.  However, for convenience we adopt the language of quantum control throughout this paper. \\
The natural measurement basis for the coupled clock system in CCG is the $\sigma_z$ basis and following the derivation in~\cite{kafri_classical_2014},  we find the master equation that describes the interaction in Eq.~\eqref{eq:HI} in CCG is
\begin{eqnarray}
\dot{\rho} \!&=&\!-\frac{i}{\hbar}[H_0 + \hbar g_{12} \sigma_z^{(1)} \sigma_z^{(2)}\!\!, \rho] - \Br{\frac{\Gamma_1}{2} + \frac{g_{12}^2}{8\Gamma_2}}[\sigma_z^{(1)}\!\!,[\sigma_z^{(1)}\!\!, \rho]] \nl
& & - \Br{\frac{\Gamma_2}{2} + \frac{g_{12}^2}{8\Gamma_1}}[\sigma_z^{(2)},[\sigma_z^{(2)}, \rho]]\,,
\label{eq:master}
\end{eqnarray}
where $g_{12} = \frac{G\hbar \omega_1\omega_2}{d_{12}c^4}$ is the Newtonian interaction rate, $\Gamma_i$ is the measurement rate of the $i^{\mathrm{th}}$ clock, and $\rho$ is the density matrix. The factor $g_{12}^2$ in the decoherence rate is due to the feedback from clock 1 onto clock 2 (and visa versa), and is required to get the correct magnitude of the $\sigma_z^{(1)}\sigma_z^{(2)}$ interaction. The double commutator term in Eq.~\eqref{eq:master} prevents entanglement of the clocks thought the $\sigma_z-\sigma_z$ interaction and leads to phase diffusion in the $\sigma_z$ basis at a rate $2g_{12}$~\cite{wiseman2009quantum}. This phase diffusion induces a fundamental limit on the time resolution of each clock that can not be avoided. Note that pure unitary evolution under the Hamiltonian in Eq.~\eqref{eq:HI} will also result in apparent dephasing if only a single clock is measured. Indeed this is exactly the type of decoherence considered in Ref.~\cite{castro-ruiz_entanglement_2015}. However, measurements of the full bipartite system will be able to show violations of a Bell inequality as the apparent decoherence is due to the two clocks becoming maximally entangled, and therefore the dephasing on a single clock is a second order (in time) effect~\cite{altamirano_unitarity_2016,grimmer_open_2016}. In contrast, CCG results in first order (in time) dephasing, and the entanglement forbidding LOCC nature of CCG means that Bell inequalities will always be satisfied. This difference between quantum and classical interactions, and the $1/d$ scaling of the dephasing rate may be used to distinguish CCG dephasing from other sources of quantum noise. \\

For two clocks of equal frequency, where one would expect the measurement rates to be equal by symmetry, the dephasing rate is minimized when $\Gamma_1 = \Gamma_2 = g_{12}/2$; for petahertz clocks ($\omega_1 = \omega_2 = 2\pi \times 10^{15}$ Hz) as used in~\cite{chou_frequency_2010,rosenband_frequency_2008} separated by 300 nm, the dephasing rate is $g_{12}/2\approx 10^{-42}$ Hz. Such a small rate would require a clock with fractional uncertainty below $10^{-57}$ to observe, and therefore cannot be ruled out by current state of the art atomic and ion clocks which have achieved a fractional uncertainty of $10^{-18}$~\cite{chou_frequency_2010,hinkley_atomic_2013,bloom_optical_2014}. \\

\begin{figure}
\def\svgwidth{\columnwidth}\scriptsize
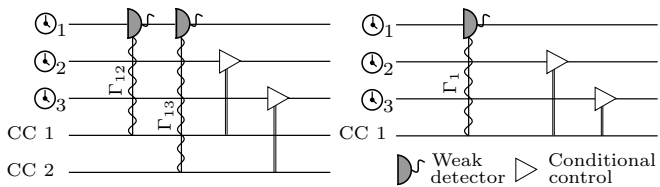
\caption{\label{fig:1}Different interpretations of the classical channel structure in CCG. We have shown only the measurements of a single clock for simplicity, but it is understood that each clock is treated equally. (left) A unique channel between each of the pairwise coupled quantum degrees of freedom as considered in \cite{MagNaty}. (right) A single channel used for each particle used for global feedback on all other particles \cite{neutronstar}.}%
\end{figure}

\textit{Multiparticle interaction}\label{sec:multiparticle} --- We now extend the analysis to $N$ interacting clocks, and we investigate the enhancement of the dephasing rate due to the multiple (order $N^2 - N$) interactions. Before preceding we have to consider how information propagates in the classical channel model. There are two possibilities of information propagation; pairwise measurement and feedback, Fig.\ref{fig:1}(left), or single measurement with global feedback, Fig.\ref{fig:1}(right). Note that both of these models are equivalent for $N=2$ clocks and hence were not discussed in Ref.~\cite{kafri_classical_2014}. The dissipative evolution of the master equation for the pairwise measurement and feedback is 
\begin{eqnarray}
\dot{\rho} &=& -\sum_i \sum_{j\neq i}\Br{\frac{\Gamma_{ij}}{2} + \frac{g_{ij}^2}{8\Gamma_{ji}}}[\sigma_z^{(i)}, [\sigma_z^{(i)},\rho]]
\label{eq:pair}
\end{eqnarray}
where $g_{ij} = g_{ji}=\frac{G\hbar \omega_i\omega_j}{d_{ij}c^4}$ is the interaction rate between clocks $i$ and $j$ and $\Gamma_{ij}>0$ is the decoherence (dephasing) rate from the measurement of clock $i$ to generate the interaction between clocks $i$ and $j$. Note that $\Gamma_{ij}$ is related to the measurement strength, with $\Gamma_{ij} = 0$ corresponding to no measurement and $\Gamma_{ij}\rightarrow \infty$ corresponds to projective measurement of $\sigmaz$. For the moment each of the $\Gamma_{ij}$'s are still free parameters; later we show that there is a non-zero set of $\Gamma_{ij}$'s that minimize the decoherence, and thus the minimum decoherence rate has no free parameters. The decoherence terms in Eq.~\eqref{eq:pair} can be intuitively understood. Each clock  is measured $N-1$ times, and each of these measurements are used to apply a independent feedback on the other $N-1$ clocks in the system. The measurements therefore contribute to a total dephasing rate of $\sum_{j\neq i}\Gamma_{ij}/2$ on the $i^{\text{th}}$ clock. The $\sum_{j\neq i}g_{ij}^2/8\Gamma_{ji}$ term for the $i^{\text{th}}$ clock is from the feedback of the $N-1$ noisy measurements form each of the other clocks in the system. Note that similar to the two clocks case, the presence of $g_{ij}^2$ in the feedback term is necessary in order to recover the correct magnitude of the systematic gravitational interaction. 

In contrast to the pairwise measurement and feedback, the global feedback only requires a single measurement of the $i^{\text{th}}$ clock (single dephasing rate $\Gamma_i$), and the single measurement result is used to apply feedback on each of the other $N-1$ clocks. For global feedback, the dissipative evolution is given by
\begin{eqnarray}
\dot{\rho} &=& -\sum_{i}\Br{\frac{\Gamma_i}{2} + \sum_{j\neq i}\frac{g_{ij}^2}{8\Gamma_j}} [\sigma_z^{(i)}, [\sigma_z^{(i)},\rho]].
\label{eq:global}
\end{eqnarray}
The dependence of $g_{ij}$ (which in turn depends on $d_{ij}$) in the dephasing rate of Eq.~\eqref{eq:pair} and Eq.~\eqref{eq:global} means that the dephasing rate of the $i^{\text{th}}$ clock depends on the spatial arrangement of the other $N-1$ clocks. 

\renewcommand{\arraystretch}{1.8}
\begin{table}%
\centering
\begin{tabular}{c c c c}
\toprule
			& \textbf{SCENARIO}         &     &   \textbf{SCALING}  \\ \hline\hline
$A.i$ &  \textbf{Pairwise} 								&  (1D) &  $\log(N) $  \\ 
      & ${\mathcal{D}\subrm{pw}}^{\!\!\!\!\!\!\!(i)} = \frac{G\hbar \omega^2}{2c^4}\sum_{j\neq i}d_{ij}^{-1}$  &(2D) & $\sqrt{N}$ \\
			& 																& (3D) & $N^{2/3 }$ \\ \hline 
$A.ii$& \textbf{Global}                 & (1D)  & $\sqrt{1 - 2/N}$ \\
			&  ${\mathcal{D}\subrm{gl}}^{\!\!\!\!\!(i)} = \frac{G\hbar\omega^2}{2c^4}\sqrt{\sum_{j\neq i}d_{ij}^{-2}}$  & (2D) & $\sqrt{\log(N)}$ \\
			&                                 & (3D) & $N^{1/6}$    \\ \hline 
$B.i$ & \textbf{Pairwise}                 & (1D)  & $\sqrt{N}$ \\
			&  ${\mathcal{G}\subrm{pw}}^{\!\!\!\!\!\!\!(i)} = \frac{G\hbar \omega^2\sqrt{N-1}}{2c^4}\sqrt{\sum_{j\neq i}d_{ij}^{-2}}$  & (2D) & $\sqrt{N\log(N)}$ \\
			&                                 & (3D) & $N^{2/3}$    \\ \hline 
\end{tabular}
\caption{MINIMUM DEPHASING RATES. The first column presents our results on minimization for the cases where $\Gamma_{ij}$ is pairwise defined (case $A$) and  it is a fundamental constant (case $B$) as outlined in the Appendix. The second column shows the scaling with the number of clocks in the array for different dimensions, assuming $N \gg 1$ by using Eq.~\eqref{integralclocks}. The coefficient of the scaling is $G\hbar\omega^2/(2 L_c c^4)$, where $L_c$ is the characteristic separation between adjacent clocks in the lattice.}
\label{deph_table}
\end{table}

We consider $N\gg 1$ clocks in 1, 2 and 3D lattice configurations with a lattice constant $L_c$. In this case we can find the minimum dephasing rate on a single clock (see Appendix). Our results are presented in the second column of Table \ref{deph_table}. Note that there are two different scenarios we can consider for minimization. The first one (case $A$) is a pairwise minimization: the rate $\Gamma_{ij}$ depends only on the separation between the clocks $i$ and $j$, and the minimization is taken before including it in the general master equation. The second scenario (case $B$) minimizes the total noise in the master equation.  One can estimate the scaling of these rates by assuming that $L_c$ is small compared to the macroscopic length scale, $R$, of the lattice, $L_c\ll R$. In this case, the summations in Table~\ref{deph_table} are well approximated by an integral expression, Eq.~\eqref{integralclocks}, for $N \gg 1$. Note that the integral approach only differs from the analytical sum by a factor of order one. We show how the dephasing rate depends on $N$ in the last column  of Table \ref{deph_table}.
With the current experiments ($N=10^6$, $L_c \approx 800$ nm \cite{akatsuka_optical_2008}) we compute a dephasing rate of order $10^{-40}$Hz (similar for all arrangements). Note that in order to have a dephasing of the order of mHz, which can be detected in the lab,  we need to have either a large number of clocks or small separation between them. For example, as considered in \cite{castro-ruiz_entanglement_2015} taking $N=10^{23}$, $L_c=1$ fm, and a 10 GeV clock transition ($10^{26}$ Hz), results in a dephasing rate of order $1$ Hz. Note however, the $\sim 1$ Hz dephasing rate here includes the spatial distribution of the clocks, whereas the the dephasing rate quoted in~\cite{castro-ruiz_entanglement_2015} assumed a 1 fm distance between each of the $10^{23}$ clocks. More realistically, we consider the M{\"o}ssbauer effect in $^{109m}\!$Ag~\cite{bayukov_observation_2009} which has a transition frequency of $8\times 10^5$ THz and a linewidth of 10 mHz, with adjacent atoms separated by $\approx$ 1 \AA. Using these parameters, CCG would predict a minimum $\gamma$-ray line width of $0.01$ nHz per $100$ g (1 mole) of $^{109m}\!$Ag, far below current experimental precision. To observe the linewidth at the order of mHz, would require $N\approx 10^{36}$ atoms, or $10^{12}$ kg of metallic silver.

%

\textit{Clocks in earths gravitational field} \label{sec:redshift}--- In the previous sections we were only concerned with energy-energy coupling between spatially separated clocks. However, the gravitational redshift is a relativistic effect that has been detected in quantum systems~\cite{hinkley_atomic_2013,hafele_around--world_1972}, and therefore is a promising candidate to study the decoherence effects predicted by CCG. Again from the mass energy equivalence, a trapped two level system with position operator $x$ will interact with any nearby object of mass $m$, and position operator $X$ via the Newtonian interaction
\begin{eqnarray}
H_I &=& -\hbar\frac{G m \omega \sigma_z}{c^2|X - x|} \nl
 &\approx& -\hbar\frac{G m\omega}{c^2|d|}\sigma_z  +\hbar\frac{Gm  \omega }{c^2d^2}\sigma_z(\delta X - \delta x)
\end{eqnarray}
where $\delta x$ and $\delta X$ are deviations about the mean separation $d$ between the clock and the mass\footnote{Note that we have not included the $\delta X \delta x$ term considered by Ref.~\cite{kafri_classical_2014}. This term is present but appears at sub-leading orders in this description.}. The first term is the mean redshift on the clock from the presence of the rest mass, and the second term is the lowest order Newtonian interaction between the quantum degrees of freedom. The $\sigma_z\delta x$ is a \textit{local} interaction between the external and internal degrees of freedom of a single particle and therefore does not need to be mediated by a classical information channel. The $\sigma_z \delta X$ term however, is a non-local interaction and is replaced by an effective measurement and feedback process in CCG. In the following we consider the dephasing of a single clock from treating the nearby mass as both a composite and simple particle, where the simple particle case is just the $N= 1$ limit of the composite particle description. For the composite particle description, we treat each constituent atom as individual point particle contributing to the redshift. The dissipative part of the CCG evolution is 
\begin{eqnarray}
\dot{\rho}_{\mathrm{diss}} \!&=&\! - \sum_i\Br{\frac{\Gamma_i}{2} + \frac{g_i^2}{8 \Gamma_z}}[\delta X_i, [\delta X_i, \rho]] \nl
& &  -\Br{\frac{\Gamma_{z}}{2} + \sum_i \frac{g_i^2}{8\Gamma_i}}[\sigma_z, [\sigma_z, \rho]]
\label{eq:redshift}
\end{eqnarray}
where $\Gamma_i$ is now the decoherence from the measurement of the position of the $i^{\mathrm{th}}$ atom, $\Gamma_z$ is the decoherence due to measurement of the clock, and $g_i = \frac{Gm_i\omega}{c^2 d_i^2}$ is the energy-position interaction between the clock and the $i^{\mathrm{th}}$ atom of mass $m_i$ and has units of in Hz m$^{-1}$. Here we have assumed the single measurement-global feedback interpretation of the model - figure~\ref{fig:1} (right) - which was shown 
previously to result in a lower bound for the minimum decoherence rate. The double commutator in position leads to momentum diffusion (heating) of each atom, an effect that  is investigated in \cite{MagNaty,neutronstar}. This heating is not unique to CCG, and has been predicted in continuous spontaneous localization models~\cite{ghirardi_unified_1986,smirne_dissipative_2015,li_detecting_2016} and stochastic extensions to the Schrodinger-Newton equation~\cite{nimmrichter_stochastic_2015}. Eq.~\eqref{eq:redshift} shows that CCG predicts a non-zero dephasing rate that accompanies the redshift, and a finite heating rate to nearby massive particles. \\
As $g_i$ scales as $1/d^2$, the dephasing due to the $g_i^2$ term in Eq.~\eqref{eq:redshift} scales as $1/d^4$, meaning that only the closest particles to the clock significantly contribute to the dephasing rate. This is easily seen by considering a macroscopic homogenous body of $N$ atoms of equal mass $m_i = m$ (for example a single species atomic crystal) close to the clock. For such a macroscopic object, one would expect by symmetry the measurement rate of each atom to be identical, $\Gamma_i = \Gamma$. By considering gravitational interactions between neighboring atoms we use the result of Ref.~\cite{kafri_classical_2014} and find $\Gamma = Gm^2/\hbar L_c^3$ where $L_c$ is now the characteristic separation between adjacent atoms (e.g. lattice constant for a crystal). In this case we can use Eq.~\eqref{integralclocks} to express the dephasing rate as an integral over the volume $V$ of the macroscopic object,
\begin{eqnarray}
\frac{G\hbar L_c^3 \omega^2}{8 c^4}\sum_i\frac{1}{d_i^4} \approx \frac{G\hbar L_c^3 \omega^2}{8 c^4}\int_V\frac{dV}{L_c^3 |r-r_0|^4}
\end{eqnarray}
where $r_0$ is the mean location of the clock, and we have used $\Gamma = Gm^2/\hbar L_c^3$. Note that the integral must converge as the point $r = r_0$ cannot be in $V$.  This integral is non-trivial for a spherical body, nevertheless there is some intuition to be gained by considering a shell of mass centered around the clock even though there is no net redshift at the center of a mass shell. For a shell with inner radius $l$, outer radius $L$, the dephasing rate due to the redshift is given by,
\begin{eqnarray}
\mathcal{D} =\frac{\Gamma_z}{2}+ \frac{\pi G\hbar\omega^2}{2 c^4}(l^{-1} - L^{-1}).
\end{eqnarray}
Form this expression we see that it is only close by masses in a thick ($L\gg l$) shell that significantly contribute to the dephasing rate. Thus in a laboratory experiment, the dephasing will be dominated by the immediate environment of the clock, even though all particles contribute to the systematic redshift. 

Alternatively, the macroscopic particle could be treated as a single degree of freedom; the dephasing on the clock is then simply given by Eq.~\eqref{eq:redshift} with a single term in the sum, 
\begin{eqnarray}
\mathcal{D} = \frac{\Gamma_z}{2} + \frac{G^2M^2\omega^2}{8 c^4 d^4\Gamma_i}
\end{eqnarray}
where $M = Nm$ is the total mass of the macroscopic object with a single measurement rate $\Gamma_i$. For $M$ the mass of the earth and $d$ as the mean separation between the earths and clocks center of mass, we can use atomic clock experiments~\cite{bloom_optical_2014,chou_frequency_2010} to bound $\Gamma_i > 10$ Hz m$^{-2}$ and $\Gamma_z < 0.1$ mHz. These bounds are set as such experiments have not observed anomalous dephasing. From this result we conclude that any dephasing from a classical channel model of gravity would not be identifiable in any gravitational redshift measurements, and despite their precision, quantum clocks are not a desirable system to observe consequences of CCG. \\


\textit{Conclusions}  \label{sec:conclusions} --- The {\it{Classical Channel Gravity}} model proposes that the gravitational interaction between quantum systems is mediated by a classical information channel that forbids entanglement of distant particles through gravity. It has also been shown to result in decoherence that is similar to that predicted by the models of Diosi and Penrose. In this work we have studied the consequences of this model when treating time operationally, this is by using two level systems as idealized clocks than an observer must use in order to define the rate of external dynamics.   Two such clocks will couple gravitationally and in the Newtonian limit this can be understood from mass energy equivalence. In this context we derive the rate at which they will decohere under CCG, and show that the minimum rate is fixed by the post-Newtonian interaction. We have also extended this analysis to optical lattice clocks in one, two and three spatial dimensions, computing how the minimum dephasing rate scales as the number of independent two level systems in the lattice. Finally we have studied a clock coupled to the earth's gravitational field and analyzed in detail the position-spin interaction in the context of the CCG model. However, due to the asymmetry between the mass-clock system we were not able to meaningfully minimize the dephasing rate. Nevertheless, we showed that the gravitational redshift must be accompanied by some dephasing with the dominant contribution being due to close by atoms. Although the model considered in this work for clocks predict dephasing, the weakness of the gravitational interaction and the sub-linear scaling with the number of particles (Table~\ref{deph_table}) give a prediction thirty-seven orders of magnitude away from the current experiments. However, note that the dephasing rates computed in this work are the minimum and it is not clear that nature will saturate this bound. This shows that despite quantum clocks being the most precise measurement devices to date and therefore seem like a natural candidate to look for deviations of standard quantum mechanics, there are not the best devices to test the CCG model. 

Let us emphasize that the dephasing present in our model is fundamental to each clock and can not be avoided as it is a consequence of reproducing the Newtonian force using only classical information. In particular, the dephasing on one clock does not depend on the quantum state of the surrounding clocks, which is consistent with  the clocks being in a separable state. Therefore, this decoherence is to be understood as a fundamental limit to temporal resolution for any clock and can not be reduced by including measurements of other clocks.  For unitary evolution of a system under the Newtonian potential, as considered in Ref.~\cite{castro-ruiz_entanglement_2015}, the decoherence appears as a result of entanglement of a single clock with a global system, and if an observer has access to the full quantum system, there is no decoherence and therefore no limit to the temporal resolution. In contrast, each clock dephasing individually in CCG means that even access to the global quantum system is not enough to resolve time with zero uncertainty. \\

\textit{Acknowledgments} --- We would like to thank Gerard J. Milburn, and Robert B. Mann for helpful discussions throughout this project. Research at Perimeter Institute is supported by the Government of Canada through Industry Canada and by the Province of Ontario through the Ministry of Research and Innovation. Research at The University of Queensland is supported by the Australian Research Council grant CE110001013.  K.K. would like to thank the University of Waterloo and the Perimeter Institute for their hospitality during the period when this work was completed. N.A. and K.K, contributed equally to the original idea, the theoretical modeling, and preparing the manuscript. 

\bibliography{Clocks}

\appendix

\section{Dephasing Rate Minimization} \label{app:minimization}
The minimum dephasing rate for the multiparticle case cannot simply be obtained by considering only a single clock. For example the dephasing of a single clock $i$ in Eq.~\eqref{eq:global} can be zero for $\Gamma_i \rightarrow 0$, and $\Gamma_{j\neq i} \rightarrow \infty$. However, this would result in each of the other $j\neq i$ clocks dephasing to a maximally mixed state instantly. Therefore, the minimization procedure must minimize each dephasing rate simultaneously to give a physically sensible result. We therefore minimize the sum of dephasing rates with respect to each of the $\Gamma_{ij}$'s (or $\Gamma_i$'s),
\begin{eqnarray}
\DD{}{\Gamma_{kl}}  \sum_i \Sq{\sum_{j\neq i}\Br{\frac{\Gamma_{ij}}{2} + \frac{g_{ij}^2}{8\Gamma_{ji}}}} &=& 0 \\
\mbox{or~~~~}\DD{}{\Gamma_k} \sum_{i}\Sq{\frac{\Gamma_i}{2} + \sum_{j\neq i}\frac{g_{ij}^2}{8\Gamma_j}} &=& 0.
\end{eqnarray}
For the pairwise feedback, the decoherence is minimized when $\Gamma_{ij} = \Gamma_{ji} = g_{ij}/2$, while for the global feedback the decoherence is minimized when $\Gamma_i^2 = \sum_{j\neq i} g_{ij}^2/4$ leading to minimum dephasing rates of  (assuming $\omega_i=\omega_j=\omega$)
\begin{eqnarray}
{\mathcal{D}\subrm{pw}}^{\!\!\!\!\!\!\!(i)} &=& \frac{G\hbar \omega^2}{2c^4}\sum_{j\neq i}\frac{1}{d_{ij}} \\
{\mathcal{D}\subrm{gl}}^{\!\!\!\!\!(i)} &=& \frac{G\hbar\omega^2}{2c^4}\sqrt{\sum_{j\neq i}d_{ij}^{-2}}
\end{eqnarray}
for pairwise and global feedback respectively. Alternatively, the measurement rates $\Gamma_{ij}$'s could be considered as some fundamental measurement rate that does not depend on the spatial distribution of the physical system. In this case, each of the $\Gamma$'s lose their $ij$ (or $j$) dependence. Nevertheless, there is still a dephasing on the clocks that can be bounded by current experiments. For a fixed $\Gamma$, the minimum dephasing on the $i$th particle is 
\begin{eqnarray}
{\mathcal{G}\subrm{pw}}^{\!\!\!\!\!\!\!(i)}&=& \sqrt{N-1}\frac{G\hbar \omega^2}{2c^4}\sqrt{\sum_{j\neq i}d_{ij}^{-2}} \\
{\mathcal{G}\subrm{gl}}^{\!\!\!\!\!(i)}&=& \frac{G\hbar\omega^2}{2c^4}\sqrt{\sum_{j\neq i}d_{ij}^{-2}}.
\end{eqnarray}
Although the dephasing from the measurement is assumed to be fixed, the total dephasing rate still depends on the local environment of the clock due to the feedback from all other clocks. For an arbitrary spatial distribution of clocks, the summations in ${\mathcal{D}\subrm{pw}}^{\!\!\!\!\!\!\!(i)}$ and ${\mathcal{D}\subrm{gl}}^{\!\!\!\!\!(i)}$ must be computed. However, for regular arrays of clocks the summations are well approximated by integrals and can be solved to find the dependence on the number of clocks $N$, and spatial distribution. If we consider a regular array of $N$ clocks with characteristic length $L_c$ between adjacent clocks, the sum can be written as,
\beq
\sum_j\frac{1}{d_{ij}^\alpha}\approx\frac{1}{L_c^D}\int_{V_D}\frac{dV_D}{r^\alpha}=\frac{S_{D}}{L_c^D}\int_{L_c}^{R}r^{D-1-\alpha} dr\,,
\label{integralclocks}
\eeq
for a clock in the center of a D-dimensional array, e.g. linear ($1D$), circular planar ($2D$), or spherical ($3D$) lattices. The integral is over the macroscopic volume, (area in $2D$ or line in $1D$) of the array, and $S_{D} = 1,~2\pi,~4\pi$ for linear, planar and spherical geometries respectively. The integral is explicitly an approximation to the sum for the $i^{\mathrm{th}}$ clock in the center of an array of radius $R = N^{1/D}L_c$. However, by using symmetry, clocks on the sides/edge of an array would be expected to have the same scaling with $N$ (which is fixed by $D-1-\alpha$), and differ only by a constant factor of order unity. For linear arrays in 1$D$ consider the following example:
\begin{eqnarray}
\sum_{j\neq i }\frac{1}{d_{ij}} &\approx& \int_{-N_L L_c}^{-L_c}\frac{1}{|x|} \frac{dx}{a}  + \int_{L_c}^{N_R L_c} \frac{1}{|x|} \frac{dx}{a} \nl
&=& \frac{1}{L_c}\log(N_LN_R)
\end{eqnarray}
where $N_L>1$ ($N_R>1$) are the number of clocks to the left (right) of the $i$th clock. As $N_L + N_R + 1 = N$, the sum scales as $\log(N)$ regardless of the physical position of the clock in the array.

\end{document}